\documentclass[12pt,preprint,referee]{aastex}

\usepackage[usenames]{color}

\usepackage{graphics}
\usepackage{graphicx}
\usepackage{epstopdf}
\usepackage{url}
\usepackage[stable]{footmisc}

\newcommand{\be}{\begin{equation}}
\newcommand{\ee}{\end{equation}}
\newcommand{\ba}{\begin{eqnarray}}
\newcommand{\ea}{\end{eqnarray}}
\newcommand{\bc}{\begin{center}}
\newcommand{\ec}{\end{center}}

\begin{document}

\title{An energy-conserving, particle dominated, time-dependent model of 3C58 and its
observability at high-energies }
\author{Diego F. Torres\altaffilmark{1,2}, Anal\'ia N. Cillis\altaffilmark{3}, Jonatan Mart\'in Rodriguez\altaffilmark{1}}

\altaffiltext{1}{Institute of Space Sciences (IEEC-CSIC),
             Campus UAB,  Torre C5, 2a planta,
              08193 Barcelona, Spain}
\altaffiltext{2}{Instituci\'o Catalana de Recerca i Estudis Avan\c{c}ats (ICREA), 08010 Barcelona, Spain}
\altaffiltext{3}{Instituto de Astronom\'ia y F\'isica del Espacio,  Casilla de Correo 67 - Suc. 28 (C1428ZAA), Buenos Aires, Argentina}

\begin{abstract}
We present a time-dependent spectral model of the nebula 3C~58 and compare it with available data.
The model is for a leptonic nebula, in which particles are subject to synchrotron, inverse Compton, self-synchrotron Compton, adiabatic, and bremsstrahlung processes. 
We find that 3C~58 is compatible with being a particle dominated nebula, with a magnetic field of  35$\mu$G. A broken power law injection fits well the multi-frequency data, with a break energy at about 40~GeV. 
We find that 3C~58 is not expected to appear in VERITAS or MAGIC~II, unless the local IR background is a factor of $\sim 20$ off Galactic models averages. 
For cases in which the CMB dominates the inverse Compton contribution, we find that 3C~58 will not be visible either 
for the Cherenkov Telescope Array.
\end{abstract}


\section{Introduction}

3C~58 was suggested to be plausibly associated with
the 831 years-old supernova SN 1181
(see e.g., Stephenson 1971, and Stephenson \& Green  2002). 
However,
recent investigations of dynamical models for the pulsar wind nebula (PWN, Chevalier 2005)
and the velocities of both, the expansion rate of the radio nebula (Bietenholz 2006)
and the optical knots (Fesen et al. 2008) imply an age 
of several thousand years. This is closer to the characteristic age of the
pulsar in the nebula, PSR J0205+6449 (Murray et al. 2002). 
A recent rekindling of the low age has been put forward by Kothes (2010), based on a new estimation of the nebula distance.

 3C~58 has a flat-radio spectrum  
 with a spectral break between the radio and IR bands
(Green \& Scheuer 1992). 
X-ray observations reveal a non-thermal
spectrum that varies with radius,
becoming steeper towards the outer regions 
(Slane et al. 2004). 
PSR J0205+6449 is  one of
the most energetic pulsars known in the Galaxy. 
The pulsar powers a faint jet and is surrounded by a toroidal structure apparently
associated with flows downstream of the pulsar wind termination
shock (Slane et al. 2004). 
 The shell of the thermal X-ray emission that was seen in 3C 58, e.g., by Gotthelf et al. (2007),
 is smaller than the maximum extent of the PWN.
 Therefore, this emission is likely associated with supernova ejecta swept up by the
 expanding PWN rather than the original forward shock from the supernova.
The pulsar has been recently detected at high-energy gamma-rays by {\it Fermi}, but only upper limits were imposed for the nebula emission (Abdo et al. 2009). Similarly, {\it Whipple} (Hall et al. 2001), and both MAGIC (Anderhub et al. 2010)  and VERITAS (Konopelko et al. 2007) observed the nebula, but only upper limits were imposed at  TeV energies. 

3C~58 and the Crab Nebula
differ significantly both in luminosity and
size.   3C~58 is larger, but 
less luminous, e.g.,  its TeV luminosity is at least $\sim$100 (Anderhub et al. 2010), 
its X-ray luminosity is $\sim$2000 
(Torii  et al. 2000), and its radio luminosity is $\sim$10~times smaller than Crab.
The similarity is in fact coming from morphology (e.g., Slane et al. 2004).

PWN models for   3C~58 have been presented before by a few authors, e.g., Bednarek \& Bartosik (2003, 2005), Bucciantini et al. (2011), with some disparity in the results, particularly at the high-energy end of the spectrum. These studies use different assumptions for the primary particles assumed to populate the wind, and differ also in the treatment of the radiative physics. With 3C~58 being a candidate for observations in the current or forthcoming generation of Cherenkov telescopes, it is interesting to study under what conditions 3C~58 is observable at high energies.
Here we present an analysis of the nebula evolution using a detailed radiative, time-dependent, leptonic code. 

\section{The PWN model and results}

The
PWN model (see
Mart\'in, Torres \& Rea 2012) uses the
diffusion-loss equation 
$
{\partial N(\gamma,t)}/{\partial t}=-{\partial}/{\partial \gamma}\left[\dot{\gamma}(\gamma,t)N(\gamma,t) \right]-{N(\gamma,t)}/{\tau(\gamma,t)}+Q(\gamma,t).
$
%
$\dot{\gamma}(\gamma,t)$ contains
the energy losses due to  
synchrotron, (Klein-Nishina) inverse Compton, bremsstrahlung, and adiabatic expansion.
$Q(\gamma,t)$ represents the injection of particles
per unit energy and unit volume in a certain time, and
$\tau(\gamma, t)$ is the escape time (assuming Bohm diffusion).
%
The model computes the luminosity produced by 
synchrotron, inverse Compton (IC, with the cosmic-microwave background as well as with IR/optical photon fields), 
self-synchrotron Compton (SSC), and
bremsstrahlung processes, all devoid of any radiative approximations. 

We assume that the spin-down of the pulsar powers the nebula:
$
L(t)=4\pi^2 I {\dot{P}}/{P^3}=L_0 \left(1+{t}/{\tau_0} \right)^{-(n+1)/(n-1)}.
$
$P$ and $\dot{P}$ are the period and its first derivative and $I$ is the pulsar
moment of inertia. The spin-down power can also be written (see second equality above) 
in terms of the 
the initial luminosity $L_0$, the initial spin-down timescale $\tau_0$, and
the braking index $n$.
$\tau_0$ is given by (e.g., Gaensler \& Slane 2006),
$
\tau_0={P_0}/[{(n-1)\dot{P}_0}]={2\tau_c}/[{n-1}]-t_{age},
$
where
$P_0$ and $\dot{P}_0$ are the initial period and its first derivative and $\tau_c$ is the characteristic age of the pulsar, 
$
\tau_c={P}/{2\dot{P}}.
$
The braking index is unknown for the great majority of pulsars, and assumed to be
$\sim 3$ (corresponding to a dipole spin-down rotator).

We adopt a broken power-law for the injection of particles,
\begin{equation}
\label{injection}
Q(\gamma,t)=Q_0(t)\left \{
\begin{array}{ll}
\left(\frac{\gamma}{\gamma_b} \right)^{-\alpha_1}  & {\rm for }\gamma \le \gamma_b,\\
 \left(\frac{\gamma}{\gamma_b} \right)^{-\alpha_2} & {\rm for }\gamma > \gamma_b,
\end{array}  \right .
\end{equation}
where $\gamma_b$ is the break energy, the parameters $\alpha_1$  and $\alpha_2$
are the spectral indices. The maximum Lorentz factor of the particles is limited by confinement, with the Larmor radius being smaller than the termination shock, 
$\gamma_{max}(t)=({\varepsilon e \kappa}/{m_e c^2})\sqrt{\eta {L(t)}/{c}},
$
where $e$ is the electron charge and $\varepsilon$ is the fractional size of the radius of the shock. 
The Larmor Radius is
$R_{L}=(\gamma_{max} m_e c^2)/(e B_s)$, where $B_s$ is the post-shock field strength, defined as
$B_s \sim (\kappa (\eta L(t)/c)^{0.5})/R_s ,$
with $R_s$ the termination radius.
We have fixed $\kappa$, the magnetic compression
ratio, to 3 (Venter \& de Jager 2006, 
Holler et al. 2012). 
The normalization of the injection function is obtained from
$
(1-\eta)L(t)=\int_0^\infty \gamma m c^2 Q(\gamma,t) \mathrm{d}\gamma,
$
where $\eta=L_B(t)/L(t)$ is the magnetic energy fraction,  assumed constant along the evolution, with $L_B(t)$ being the magnetic power,
and $B$ is the average field in the nebula.

We have adopted a $B(t)$ resulting from 
\begin{equation}
\int_0^t{\eta L(t')R_{PWN}(t')dt'}=(4\pi/3)R_{PWN}^4(t)B^2(t)/(8\pi).
\end{equation}
This equation is equivalent to 
$(dW_B/dt)=\eta L-W_B(dR_{PWN}/dt)/R_{PWN}$
where 
$W_B=(4\pi/3)R_{PWN}^3(t)B^2(t)/(8\pi)$,
which includes the adiabatic losses due to nebular expansion (e.g., Pacini \& Salvati 1973).

We have adopted 
the free expanding expansion phase in the model by van der Swaluw et al. (2001, 2003), where the radius of the
 PWN is 
$
 R_{PWN}(t) \sim \left({L_0 t}/{E_0} \right)^{1/5} V_{ej} t,
$
with 
$V_{ej}=\sqrt{{10 E_0}/{3 M_{ej}}}$ and where 
$E_0$ and $M_{ej}$ are the energy of the supernova explosion and the ejected mass, respectively. 
 %
 %
$V_{ej}$ is determined requiring that the kinetic energy of the ejecta equals $E_0$. 
These assumptions encompass self-similarity of the ejecta flow, 
a linear velocity profile as well as a uniform density ejecta, and it is
the same to that taken by Bucciantini et al. (2011) with their density distribution parameter $\alpha=0$.
Adding more freedom to the dynamics by changing $\alpha$ does not have a significant impact in the results.
We adopt a mass of the ejecta comparable to that of Crab, motivated by estimates of the total mass of the precursor (Rudie \& Fesen 2007).


\begin{table}[t!]
\centering
  \caption{Physical magnitudes}
  \begin{tabular}{@{}lll@{}}
  \hline
  From observations, computed, or assumed \\
  \hline
  \hline
   Magnitude & Value    \\
 \hline
  \hline
 Period (ms), $P(t_{age})$         & 65.7  \\
 Period derivative (s s$^{-1}$), $\dot{P} (t_{age})$& 1.93$\times 10^{-13}$  \\
 Characteristic Age (yr),                  $\tau_{c}$           & 5397 \\
 Moment of inertia (g cm$^{2}$),           $I$                  & $10^{45}$ \\
 Spin-down luminosity now (erg/s),         $L(t_{age})$         & $2.7\times10^{37}$ \\
 Braking index,                           $n$               & 3        \\
 SN explosion energy (erg),                $E_{0}$              & $10^{51}$\\
 Ejected mass ($M_{\odot}$),               $M_{ej}$             &   8      \\
 Hydrogen density (cm$^{-3}$),             $n_{H}$              &   0.1 \\
\hline
Adopted age and distance, derived quantities \\
\hline
 Age (yr),                                 $t_{age}$   & 2500   \\        
 Distance (kpc),                           $d$         & 3.2                \\             
 Initial spin-down age (yr),               $\tau_{0}$  & 2897     \\           
 Initial spin-down luminosity (erg/s),     $L_{0}$     & $9.32\times10^{37}$ \\ 
\hline
\end{tabular}
\label{parameters}
\end{table}

\begin{figure}[t!]
\centering
\includegraphics[angle=0, scale=0.45] {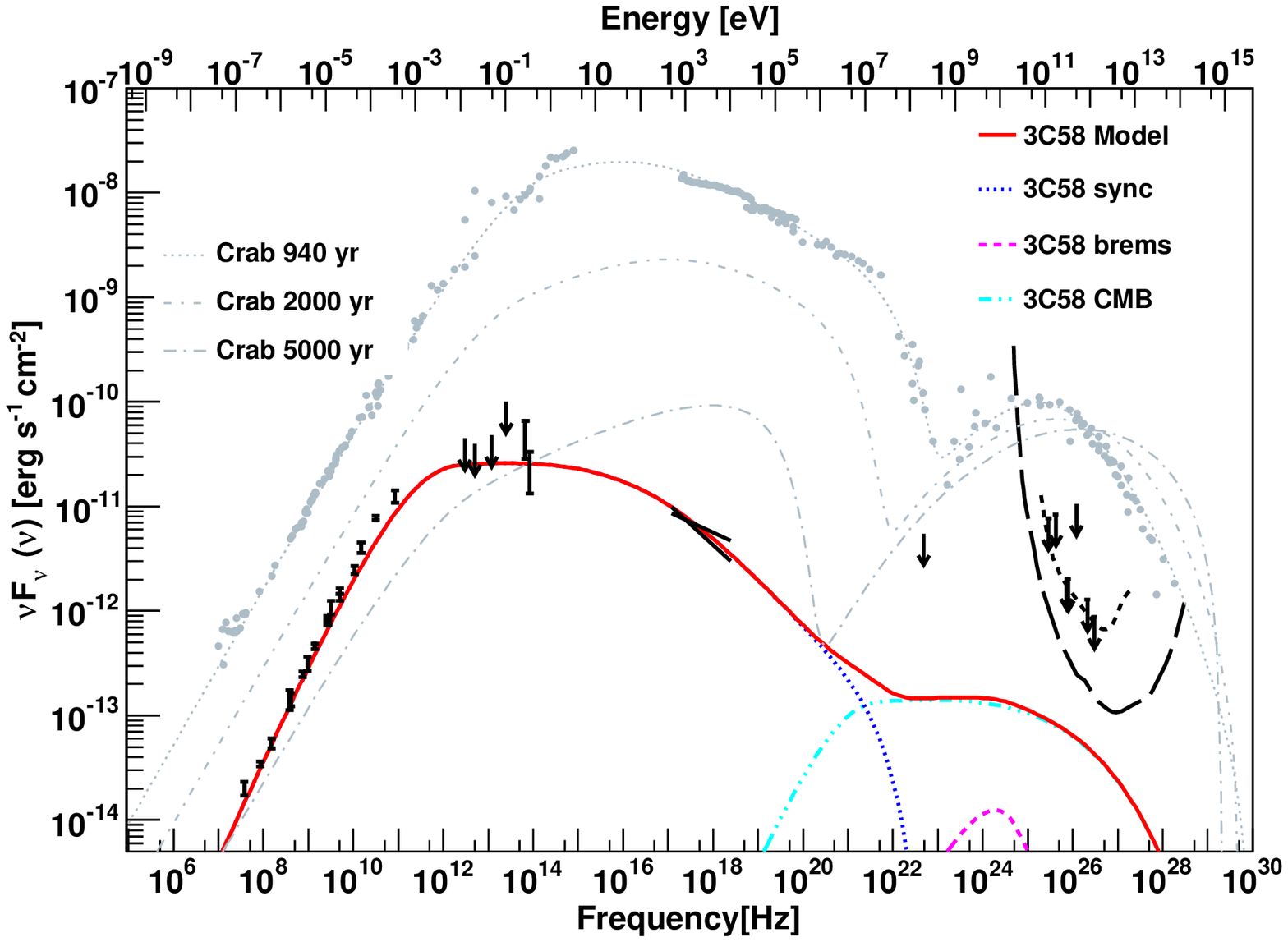}
\includegraphics[angle=0, scale=0.45] {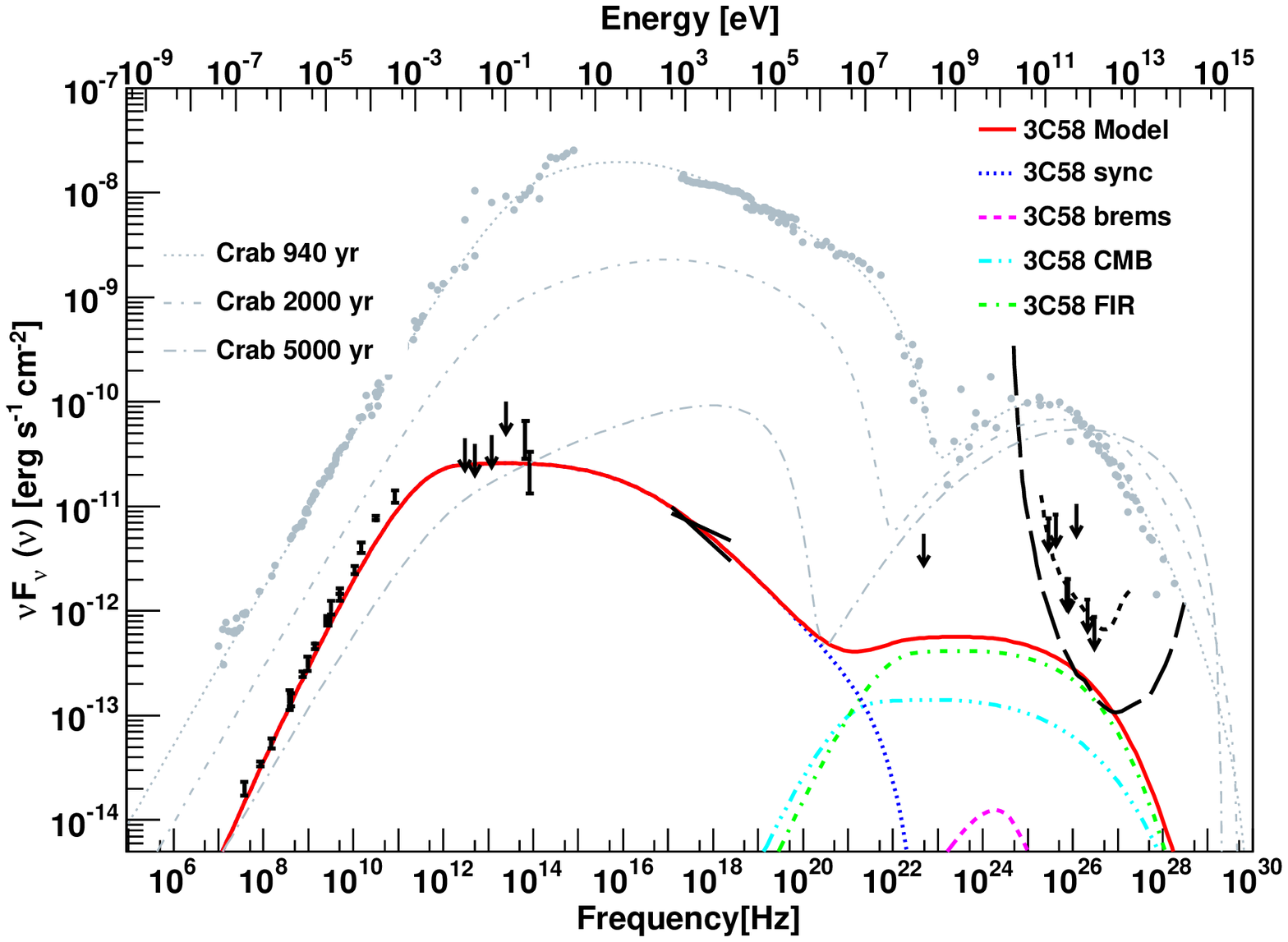}
\includegraphics[angle=0, scale=0.45] {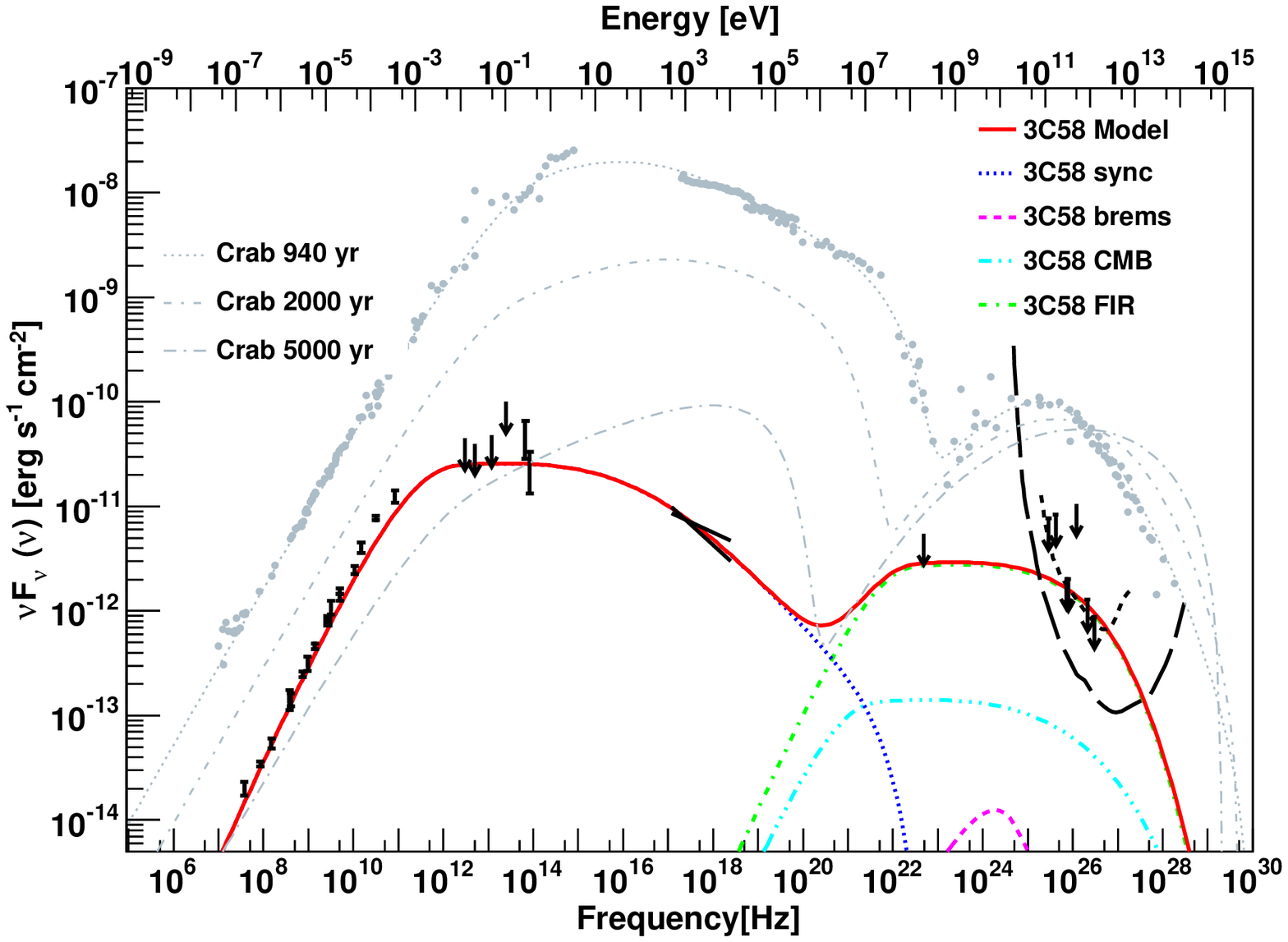}
\caption{Multi-frequency models of the PWN   3C~58 under different assumptions for the background
photon fields. Top: CMB-only. Middle: IR energy density up to the level where the emission of   3C~58 reaches the
CTA sensitivity. Bottom: {\it ibid}, for the MAGIC II sensitivity. Fitting parameters are giving in Table \ref{fit}. The SSC contribution is not visible in this scale. }
\label{MW-2500}
\end{figure}

\begin{table}
 \centering
  \caption{Fit values}
 \begin{tabular}{@{}llll@{}}
    \hline
Without IR contribution to IC \\
  \hline
  \hline
   Magnitude & Value                 \\
  \hline
  \hline
 Max. Lorentz factor today, $\gamma_{max} (t_{age})$ & $7.3\times10^9$ \\
 Break Lorentz factor,          $\gamma_{b}$                      & $0.78\times 10^5$    \\
 Low energy index,                         $\alpha_{1}$              & $1.05$               \\
 High energy index,                        $\alpha_{2}$              & $2.91$                 \\
 Shock radius fraction,                    $\epsilon$                 & $0.3$                \\
 Magnetic field today ($\mu$G),    $B(t_{age})$              & $35$           \\
 Magnetic fraction,                           $\eta$                         & $0.21$               \\
 PWN radius today (pc),         $R_{PWN}(t_{age})$       & $3.7$           \\
  \hline
  \hline
IR contribution up to CTA sensitivity \\
    \hline
   IR temperature (K),                      $T_{FIR}$                 &  $ 20$          \\
 IR energy density (eV/cm$^{3}$),         $w_{FIR}$                 &  $ 0.75$         \\ 
  \hline
  \hline
IR contribution up to MAGIC-II sensitivity \\
    \hline
   IR temperature (K),                      $T_{FIR}$                 &  $ 20$          \\
 IR energy density (eV/cm$^{3}$),         $w_{FIR}$                 &  $ 5.0$         \\ 
\hline
\end{tabular}
\label{fit}
\end{table}
 
Not all estimates 
for the age and distance of   3C~58 (see Fesen et al. 2008 for a summary) can be consistently encompassed within
the expansion model of the nebula. 
Consider first an age of $\sim$5000 years or more 
(as in Murray et al. 2002, Bietenholz 2006, Slane et al. 2002) and a distance of 3.2 kpc (as in Roberts et al. 1993).
Following Fesen et al. (2008) for the angular size of the nebula, its physical size at that distance is $6 \times 9.5$ pc. We can extrapolate 
these magnitudes to the spherical case by matching the
projected area of the nebula to that of a circle, and so we obtain a radius of 3.7 pc. Using $P$ and $\dot P$,  the parameters
of the first panel of Table \ref{parameters}, and the formulae above, 
we would however obtain a physical size of about 18~pc; a result that worsens for larger ages. 
On the contrary, if we assume the observed size and compute the ejected mass needed to have a radius of $\sim 3.7$ pc, we would find an inconsistently large value. Similarly problematic results the scenario when we change the initial spin-down power, since it would be impossible to reach the current $L(t)$, being the initial one smaller than the current power. 
Such innuendos are not solved by assuming a different 
value of braking index, and are also stable (producing sizes in excess of 10 pc) for up to one order of magnitude variations in $I$. 
If 3C~58 is closer to Earth, the mismatch would be larger, given that the physical size of the nebula would be smaller than at 3.2 kpc.

Consider next a 830 years-old nebula (as in Stephenson 1971, 
and Stephenson \& Green 2002) and a distance of 2 kpc (as in Kothes 2010).
Geometry implies that the physical size of the nebula should be around 2.3 pc, 
but using the observed, derived, and assumed parameters of Table \ref{parameters}
we obtain a size of 0.8 pc, a factor of 3 smaller. 
This result, similarly to the larger age case above, is stable against
changes in $n$, $I$, or other parameters. The only way to recover a larger nebula would be to assume
a mass of the ejecta of the order of 1 M$_\odot$, but this would be inconsistent with estimates based 
on the observed filamentary knots (which account already for a large fraction of 1 M$_\odot$) or with evolutionary models (e.g., Rudie \& Fesen 2007, Fesen et al. 2008, Bocchino et al. 2001, Slane et al. 2004).
A larger distance to   3C~58 would imply a larger physical size of the nebula, making the mismatch more severe.

We are a priori favorable to the case of an age of 2500 years 
and a distance of 3.2 kpc.
For this set of parameters, the size of the nebula can be easily accommodated within the model described in the previous section. 
Variations of the parameters in a reasonable manner maintain this conclusion stable. In addition, the shock velocity 
agrees with estimates coming from the thermal X-ray emission (e.g., Bocchino et al. 2001) and at
the same time, the swept-up mass resulting from these model parameters
$M_{sw} = M_{ej} (R_{PWN}/V_{ej}t)^3 \sim 0.26 M_\odot$ is in line with the measurements of the mass contained in filaments (Bocchino et al. 2001, Slane et al. 2004); i.e., we assume that the filamentary structure roughly corresponds to the swept-up shell of the ejecta.
Similar conclusions have been reached by Chevalier (2004, 2005) or Bucciantini et al. (2011) with others arguments.

Multi-frequency results of our model for a   3C~58 age of 
2500 years located at 3.2 kpc are shown in Fig.  \ref{MW-2500}.\footnote{The data for 3C~58 in that Figure comes
from
Green (1986), Morsi and Reich (1987), Salter et al. (1989) [in radio],
Green (1994), Slane et al. (2008) [in IR],
Torii et al. (2000) [in X-rays],
Abdo et al. (2009) [in GeV],
and Hall et al. (2001), Konopelko (2007)
and Anderhub et al. (2010) [in TeV]. Data for Crab have been collected by Martin et al. (2012).}
We also show for comparison the results corresponding to the Crab Nebula at different ages, i.e., 940 (where the data points are fit with a corresponding model, see Mart\'in et al. 2012), 2000, and 5000 years. The differences in the data of Crab and that of   3C~58 are evident.
Fig.  \ref{MW-2500} 
shows results for three different assumptions regarding the dominance of the IC contribution. In the first panel, only the CMB is assumed as background for IC. The resulting parameters
are given in Table \ref{fit}, and they show that a broken power-law fits the current radio to X-ray data. 
The magnetic field of the nebula results in $35$ $\mu$G. The contribution of SSC to IC is sub-dominant to bremsstrahlung
under the assumption of a low medium density of 0.1 cm$^{-3}$. As there is no clearly detected SNR shell, we cannot reliably estimate the ISM density. 
It is impressive how low the prediction in the GeV and TeV regimes is, far beyond the reach of {\it Fermi}-LAT 
and Cherenkov telescopes. 

Taking into account that the IR/FIR background is uncertain, we consider two limiting situations by exploring how large the energy density at 20 K should be for its corresponding IC contribution to reach the sensitivity of the Cherenkov Telescope Array (Actis et al. 2011) and of the currently operating MAGIC II (Aleksic et al. 2012). The additional parameters shown in the last two panels of Table \ref{fit} are used in order to achieve the fits shown in the middle and bottom plots of Fig.  \ref{MW-2500}. They represent energy densities ranging from 3 to 20 times  that of the CMB. These can be compared with the typical background at that frequency from e.g., GALPROP models (see Fig.  1 of Porter et al. 2006), to see that the case where   3C~58 appears as an in-principle-observable nebula in MAGIC II or VERITAS would be highly unexpected. 
In both of these cases, still, the GeV emission would be far below the {\it Fermi}-LAT upper limit.

\begin{figure}
\centering
\includegraphics[angle=0, scale=0.45] {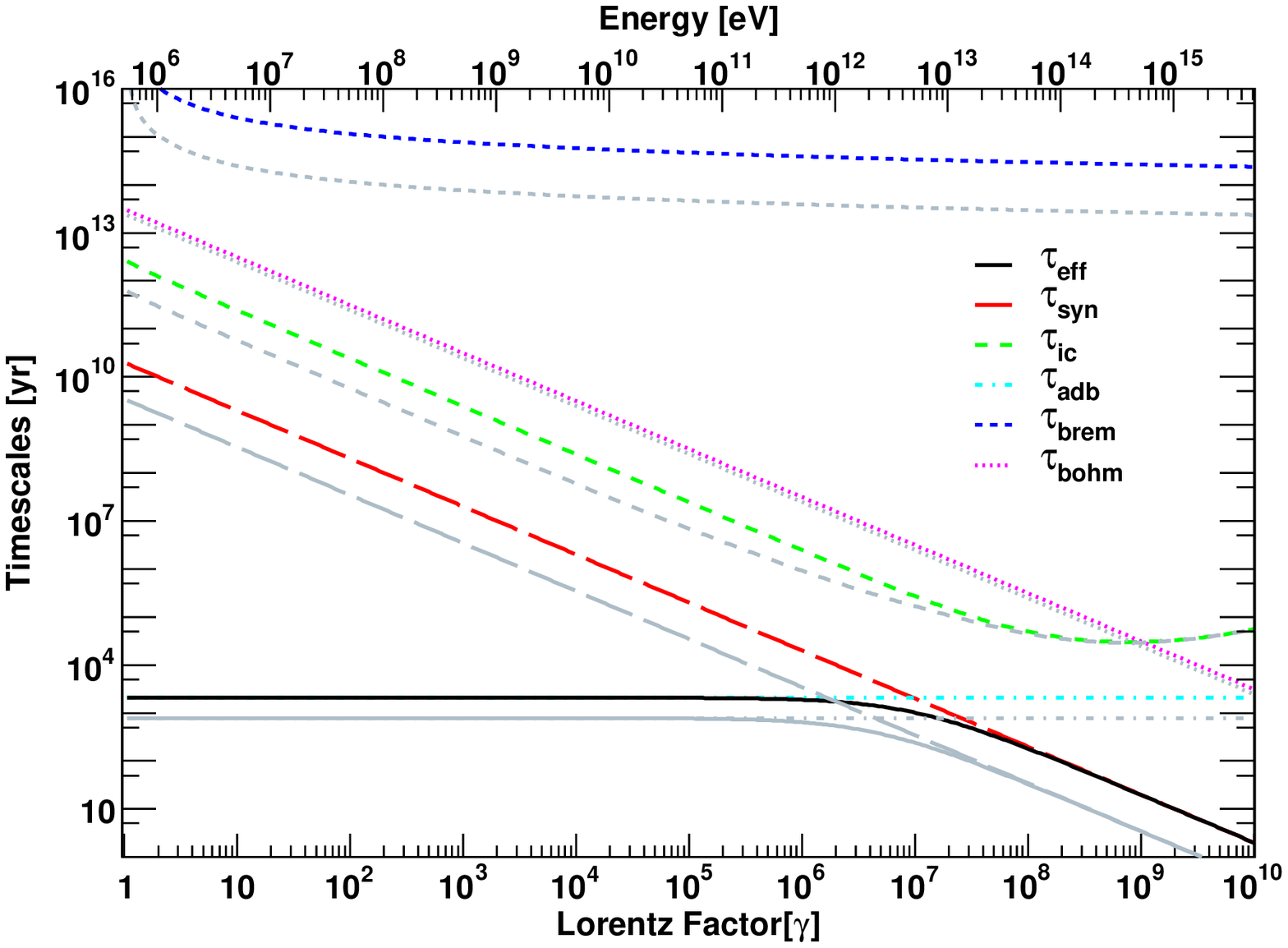}
\includegraphics[angle=0, scale=0.45] {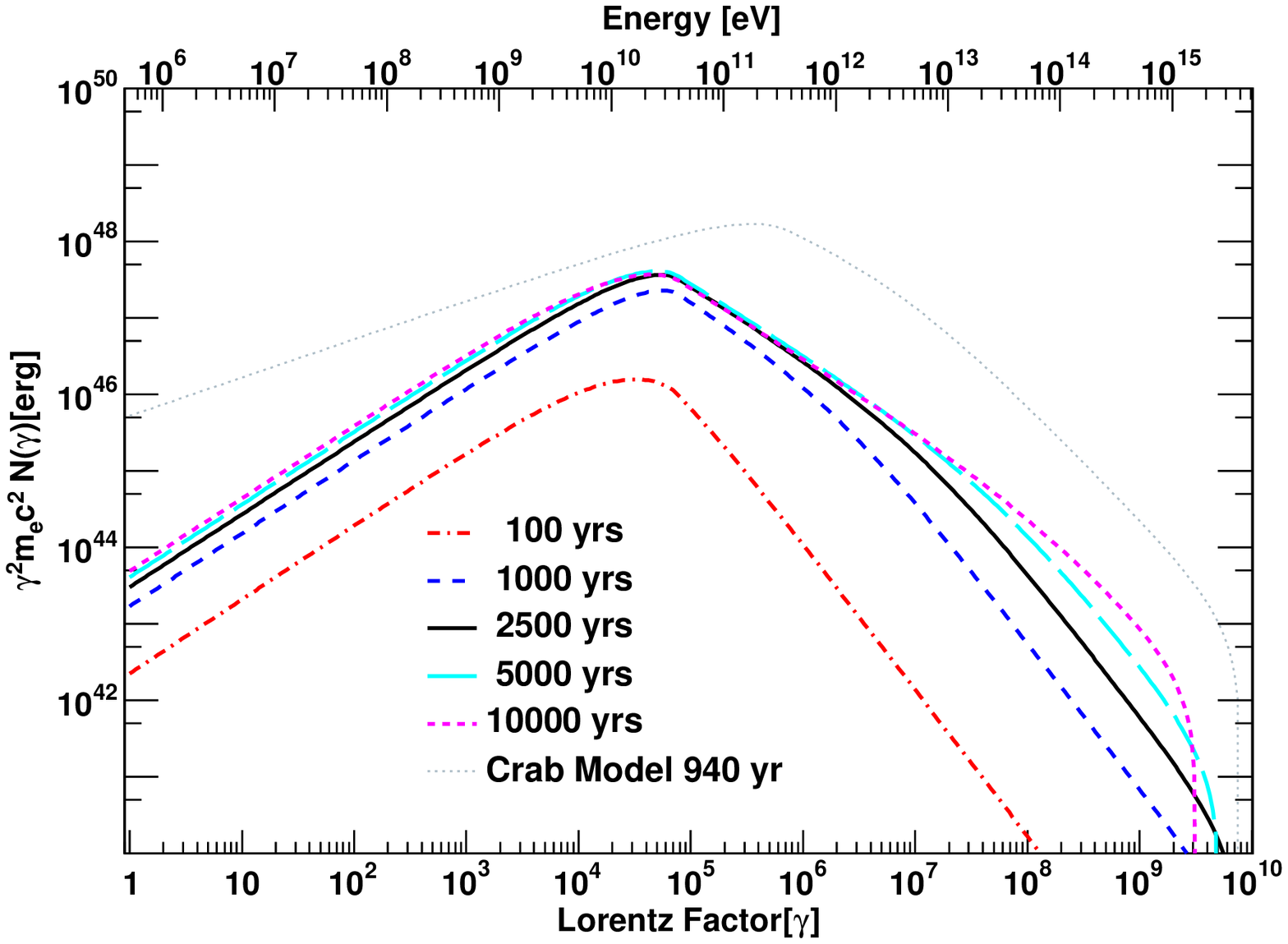}
\includegraphics[angle=0, scale=0.45] {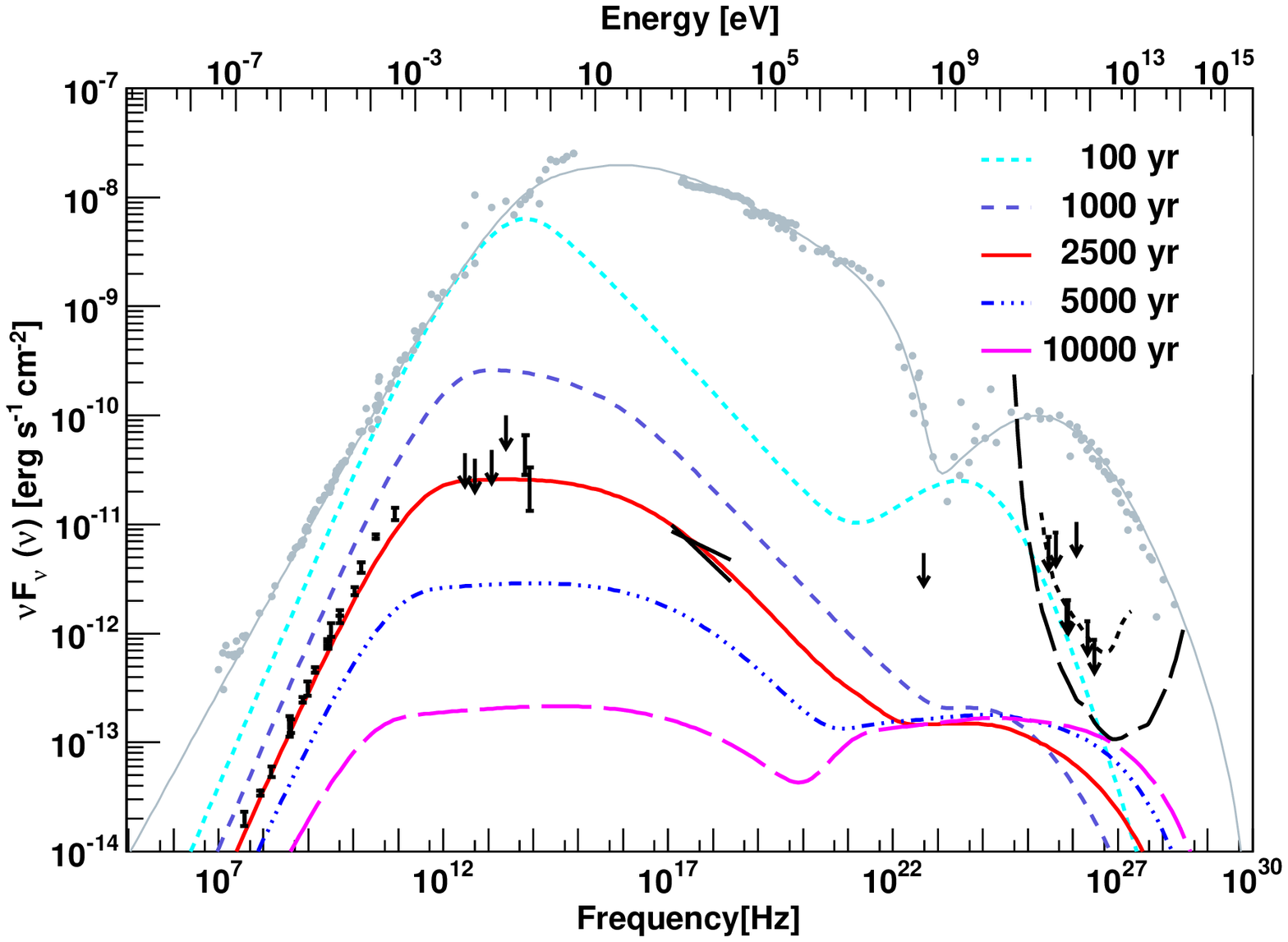}
\caption{CMB dominated model for the   3C~58 multi-frequency emission. The first panel shows the timescales for the losses under the different processes at the current age. The second panel shows the evolution of the electron spectra in time, whereas the third panel does it with the SED. Corresponding results for the Crab Nebula at its current age are shown in grey.
}
\label{MW-2500-2}
\end{figure}



\section{Discussion}

Fig.  \ref{MW-2500-2} focuses on the CMB-only model. The first panel shows the timescales for the losses at the current adopted age, showing a clear domination of synchrotron losses at high, and adiabatic losses at low energies. This is a generic feature of all models. Losses results comparable to those in the Crab Nebula today. The second panel shows the evolution of the electron spectra in time, whereas the third panel does it with the SED. The electron distribution is compared with that of Crab today, and it can be seen both a change in the peak as well as in normalization (differences amounting between 1 and 2 orders of magnitude across all the energies) between the two nebulae.

We
have explored models having different braking indices (keeping age and
distance fixed at 2500 years and 3.2 kpc, respectively, as the
testbed).
For these models, the overall quality of the fit
is unchanged; some of the fit parameters are slightly modified, however.
Lower values of $n$ implies changes in the initial spin-down power
(from 9.3, to 7.3, to 5.9 $\times 10^{37}$ erg s$^{-1}$ for $n=3$,
2.5, and 2, respectively), 
initial spin-down age (from 2897, to 4696 to 8294 years) and the PWN radius today (from 3.7 to 3.5 to 3.4 pc).
The magnetic fraction changes from 0.23, for $n=2$ to 0.22 for  $n=2.5$, to 0.21 for $n=3$. The magnetic field today
has a value between 35 and  37 $\mu$G in all these cases. 
%
Letting the $\epsilon$ parameter vary allows to explore the ability of
the fit to adapt to the
X-ray measured spectra better. But again we find a very small
parameter dependence. For instance,
for $\epsilon=0.5, 0.3,$ and 0.1, we find the maximum Lorentz factor
changing from 10.2, to 7.3,
to 2.5 $\times$ 10$^9$, with the high energy index, magnetic field,
and magnetic fraction being practically unchanged to attain fits of
similar quality of those in Fig. \ref{MW-2500}.
None of these models increment significantly the high-energy yield of
the nebula.

The total energetics is conserved in our model in exact terms, since  particles have a fraction (1-$\eta$), and the magnetic field a fraction $\eta$, of the total power. 
3C~58  features a 21\% magnetic fraction in our model, significantly higher than the one we obtain for Crab ($\sim 3\%$).
It is still a particle-dominated nebula.
These results differ from those in Bucciantini et al.'s (2011) work, where the total energy was not conserved by $\sim$30\%, leading to a nebula in equipartition,
and where it was said that models with energy conservation would always under predict the radio flux.

Note the two contiguous IR measurements around $10^{14}$~GHz. Because of their location, it is nearly impossible to fit them both at
once. Indeed, there appears
to be two subsequent steepening of the spectrum, one just beyond the
radio band and
one additional in the IR band (see Slane et al. 2008). 
We have
explored an injection containing 
another break,
but results do not improve the fit significantly. We also tried to improve the fit using an injection model based on
the PIC simulations done by Spitkovsky (2008) keeping the ratios of the additional 
parameters as in Holler et al. (2012). These are not devoid of significant extrapolations 
and include a number of additional parameters for which we have no constraints.
We used a value for the energy break of
$\gamma_b=2 \times 10^4$ to have an acceptable fit of the radio points, but it is
not possible to fit the IR and X-ray points correctly, even changing the ratios of
the parameters.
In all these cases for the injection, the fits are of similar (2 breaks power-law) or lower  (PIC motivated) quality than the ones presented above; and in none, the high-energy yield is significantly affected.

Observations at the smallest angular resolution make these studies preliminary to a yet-lacking 3D MHD/radiative/time-evolving code which could at once deal with morphology and spectral evolution. Another caveat may reside in that 3C~58 nebula is assumed to be in the free expansion phase: Green (1987) reported an increase in 3C~58 flux density 
at radio frequencies between 1967 and 1986, which might be the result of the reverse shock  encountering the PWN shock around 3C58, with the PWN being compressed and therefore the flux density raising. This report has not, however, been subsequently confirmed (Bietenholz 2006).

\acknowledgements

Work done in the framework of  the
grants AYA2009-07391, AYA2012-39303,
SGR2009-811, TW2010005 and iLINK2011-0303. 
We thank N. Bucciantini and an anonymous referee for comments and discussions.

\end{document}